\documentclass[aps,superscriptaddress,prd,
onecolumn,
floatfix, 
nofootinbib,
amsmath,amssymb,amsfonts,longbibliography]{revtex4-2}
\usepackage[pdftex]{graphicx}
\usepackage[colorlinks,linkcolor=blue,anchorcolor=violet,citecolor=red]{hyperref}

\usepackage{float,color}
\usepackage{rsfso}
\usepackage{physics}
\newcommand{\ep}{\epsilon}
\newcommand{\al}{\alpha}
\newcommand{\pa}{\partial} 
\newcommand{\del}{\delta}

\date{March 25, 2023} % revision
\begin{document}
\title{Analog de Sitter universe in quantum Hall systems with an
  expanding edge}
\author{Yasusada Nambu}
\email{nambu@gravity.phys.nagoya-u.ac.jp}
%\author{Yuki Osawa}
%\email{osawa.yuki.e8@s.mail.nagoya-u.ac.jp}
\affiliation{Department of Physics, Graduate School of Science, Nagoya
  University, Chikusa, Nagoya 464-8602, Japan}
\author{Masahiro Hotta}
\email{hotta@tuhep.phys.tohoku.ac.jp}
\affiliation{Department of Physics, Tohoku University, Sendai 980-8578, Japan}
%%%
\begin{abstract} 
  Expanding edges in quantum Hall systems can become a simulator of
  quantum 1+1 dimensional expanding universes. In these systems, edge
  excitations are represented as a chiral scalar field in curved
  spacetimes. We investigate Hawking radiation and entanglement
  behavior predicted by this model assuming that the expansion law of the edge
  region corresponds to a de Sitter universe.  As observable
  quantities for the quantum field, local spatial modes associated
  with detection regions are introduced using window functions for the
  field, and their correlations are evaluated. We found impact of
  Hawking radiation caused by the edge expansion on autocorrelation
  functions of the local modes, and confirmed that entanglement death
  due to Hawking radiation occurs. This behavior of entanglement is
  related to ``quantum to classical transition" in cosmic inflations.
\end{abstract}  
%%%%%%%%%%%%%%%%%%%%%%%%%%%%%%%%%%%%%%%%%%%%
 
%\bibliographystyle{mybib}

\maketitle
%\tableofcontents
%%%%%%%%%%%%%%%%%%%%%%%%%%%%%%%%%%%%%%%%%%%%%%%%%%% 
\section{Introduction}
Analog models of gravitational systems are useful to understand the
physics of black holes and early universe \cite{Barcelo2005}. Indeed,
although quantum aspects of black hole evaporation with Hawking
radiation has been paid much attention to theoretically, it is
difficult to observe the phenomena in our universe due to its too low
temperature for astrophysical black holes. Concerning cosmological
particle creations, it is hard to observe the occurrence of this effect
in the early stage of the universe directly. However, considering
condensed matter systems, it is possible to design models with causal
horizons for excitations, which have similar properties as black
hole horizons or cosmological horizons for null rays in general
relativity.  Thus it is possible to perform experiments of analog
Hawking radiation \cite{Steinhauer2016a,MunozdeNova2019} and particle
creations in expanding universes in laboratories.  Quantum Hall (QH)
systems can become one of such analog models
\cite{Hegde2019,Dalui2020,Subramanyan2021}. A QH system emerges when a
strong perpendicular magnetic field is applied to two-dimensional
electrons when the Landau level filling factor becomes an integer or a
rational fraction \cite{Yoshioka2002, Tong2016}. The QH systems are
typical topological materials consisting of the bulk and edge. In the
bulk, the dynamics yields a large energy gap in its dispersion
relation. In the edge, the dispersion relation of the edge current is
protected owing to the topological structure of the system, and the
edge excitations are always gapless. Thus the edge effective theories
are given by free field theories with a chiral condition, and belong
to a class of conformal field theory in 1+1 dimensional spacetime.

Most of all experiments of QH systems have been performed in a static
situation. The electrons are confined in the bulk region by a static
electric field created by the surface potential of host semiconductors
of the 2D electrons and the edge attached to the bulk remains
unchanged in time.  Expanding edges were proposed in \cite{Hotta2014a}
and experimental realization is ongoing \cite{Kamiyama2022a,Kamiyama2022b}; the
edge expands by gradually relaxing the external electric fields
through continuous electron supply into the bulk and the excitations
moving along the edge are affected by the expansion.  Thus, it is
possible to perform experiments of the quantum scalar field in an
analog expanding universe. In our previous paper \cite{Hotta2022b}, we
formulated a quantum field theory in 1+1 dimensional curved
spacetime to analyze the edge dynamics. It was shown that the
expanding edges can be regarded as expanding universe simulators of
two-dimensional dilaton-gravity models, and pointed out that our
theoretical setup can simulate the classical counterpart of an analog Hawking
radiation with  Gibbons-Hawking temperature from the future de
Sitter horizon formed in the expanding edge region.

In this paper, applying the formulation developed in our previous
paper \cite{Hotta2022b}, we investigate quantum aspects of the scalar
field in an expanding edge model which reproduces an analog de Sitter
universe. In particular, we focus on particle creation in the analog
de Sitter universe (Hawking radiation), generation of quantum
fluctuations by edge expansion, and their entanglement behavior.  We
will show that the thermal radiation with the Gibbons-Hawking
temperature from the expanding edge region is created and that it is
detectable in a static edge region. For this purpose, instead of
introducing a specific detector model to measure the Hawking
radiation, we define local spatial modes of the scalar field using
window functions and consider their correlations. Furthermore, we will
also investigate the entanglement between two spatial regions and show
that it decreases by Hawking radiation coming from the expanding edge
region. We regard this behavior as a feature corresponding to the
disappearance of quantumness of the primordial quantum fluctuations
expected in cosmic inflations \cite{Nambu2008,Nambu2011,
  Matsumura2018}. Thus, our analog de Sitter model can be available as a
simulator of the early universe to explore the generation mechanism and
the feature of primordial quantum fluctuations originated by cosmic
inflations.

The plan of the paper is as follows. In Sec.II, we review our setup of
the expanding edge of the QH systems and an analog de Sitter universe. In
Sec. III, we present the behavior of classical wave propagation in the expanding edge
system. In Sec.IV, we formulate the quantum treatment of edge excitations
and investigate the behavior of spatial local modes which are
measurable in an experiment of the QH system. In Sec. V, we discuss entanglement between spatial
modes. Section VI is devoted to summary and speculation.
%The unit of$\hbar=1$ was adopted throughout the study.

%%%%%%%%%%%%%%%%%%%%%%%%%%%%%%%%%%%%%%%%%%%%%%%%%%%
\section{Expanding edge of quantum Hall system}

Let us consider a massless scalar field $\varphi$ on the edge of QH
systems. Based on the effective theory of the edge excitations in QH
systems \cite{Yoshioka2002,Tong2016}, the edge mode is
represented by a massless scalar field $\varphi$ the  wavelength
of which is 100 times larger than the magnetic length
%%%
\begin{equation}
    \ell_B=\sqrt{\frac{\hbar}{eB}},
\end{equation}
%%%
where $B$ is a perpendicular magnetic field.  The edge current and the
edge charge density are given as derivatives of the scalar field. We
derive the  wave equation for $\varphi$. The left moving modes and the
right moving modes of $\varphi$ obey
%%%
\begin{equation}
  \pa_\tau\varphi_L-\frac{v}{a(\tau)}\pa_x\varphi_L=0,\quad
  \pa_\tau\varphi_R+\frac{v}{a(\tau)}\pa_x\varphi_R=0,
\end{equation}
where $\tau$ denotes a time variable in a laboratory, $x$ is the
comoving coordinate along the edge and the proper length along the
edge is given by $a(\tau)\int dx$. The scale factor $a(\tau)$
represents the expansion of the edge. Using the trapping potential
$U(y)$ perpendicular to the edges of the QH system, the propagation
speed of the edge excitation $v$ is determined as
%%%
\begin{equation}
v=\frac{c\,U'(y)}{eB}=\frac{cE}{B},
\end{equation}
%%%
where $E$ is the electric field induced by $U$. This propagation speed
of the edge excitation is the same as the classical drift velocity of
electrons.  The solution of these equations is
%%%
\begin{equation}
  \varphi_L=A\left(v\int\frac{d\tau}{a}+x\right),\quad
  \varphi_R=B\left(v\int\frac{d\tau}{a}-x\right),
\end{equation}
%%%
where $A,B$ are arbitrary functions. The scalar field
$\varphi:=\varphi_L+\varphi_R$ obeys
%%%
\begin{equation}
  \ddot\varphi+\frac{\dot
    a}{a}\dot\varphi-\frac{v^2}{a^2}\pa_x^2\varphi=0,\quad\dot{}=\frac{\pa}{\pa\tau}. 
\end{equation}
%%%
This is the Klein-Gordon equation $\square\varphi=0$ in a 
1+1 dimensional expanding universe the metric  of which  is given by
%%%
\begin{equation}
  ds^2=-v^2d\tau^2+a^2(\tau)dx^2.
  \label{eq:metric1}
\end{equation}
%%%
The propagation speed of the edge excitation $v$ plays the same role
as the speed of light $c$ in general relativity which determines causal
structures of spacetimes. It is possible to control the expansion law
$a(\tau)$ by tuning the external trapping electric field for the edge
region. We can perform experiments of quantum physics of an early
universe using the analog expanding universe by analyzing QH systems
with an expanding edge.  From now on, we set $v=1$ and we use $v$ as a
unit of length and time in our analog spacetimes.  By introducing the
conformal time $t:=\int d\tau/a$ and null coordinates $x^\pm:=t\pm x$,
the metric is written as the conformally flat form:
%%%
\begin{equation}
  ds^2=-a^2 dx^+dx^-.
  \label{eq:metric2}
\end{equation}
%%%
The scalar field is represented as
%%%
\begin{equation}
  \varphi=\varphi_L(x^+)+\varphi_R(x^-).
\end{equation}
%%%
For a given form of $a(\tau)$ which represents the expansion law of the
edge region, it
is possible to identify  a corresponding  analog universe using the
metric \eqref{eq:metric2}. In the QH systems, either $\varphi_L$ or
$\varphi_R$ is allowed due to the boundary condition of QH systems.

\begin{figure}[H]
  \centering
   \includegraphics[width=0.9\linewidth]{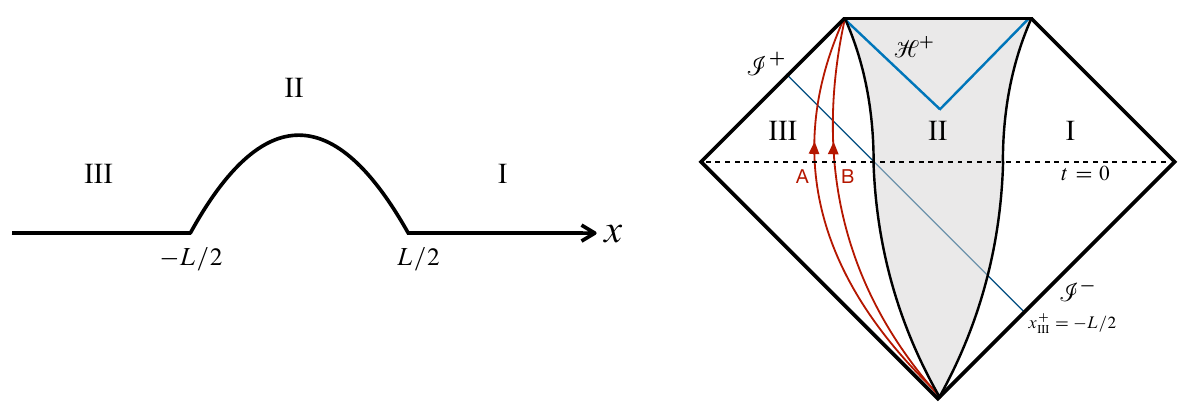}
   \caption{Left: schematic picture of the expanding edge system. $x$
     denotes a coordinate along the edge of the QH system.  Regions I and
     III are static Minkowski regions and the expanding region II
     corresponds to a de Sitter universe. Right panel: Penrose diagram
     representing the present setup with the expanding edge region II (gray region), which starts
     accelerated expansion at $t=0$. A and B denote world lines of
     detectors which perform measurements of edge excitations. This
     diagram corresponds to the $\pi/4<LH<\pi/2$ case. For parameter
     values not included in this range, global structure of the
     spacetime becomes different (see \cite{Hotta2022b}). }
  \label{fig:penrose1} 
\end{figure}

\begin{figure}[H]
  \centering
  \includegraphics[width=0.9\linewidth]{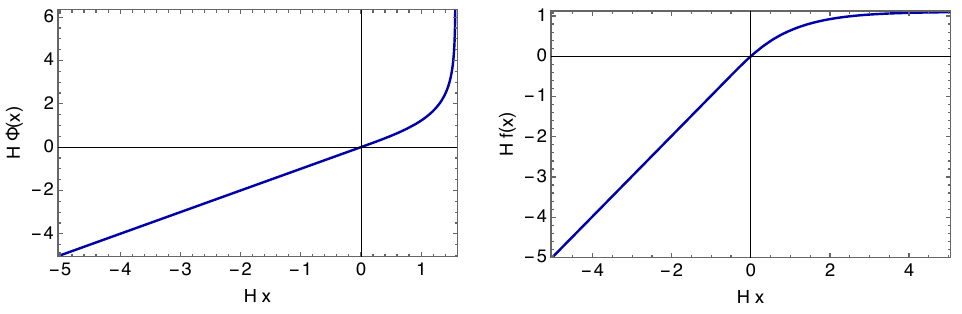}
  \caption{Left panel: the function $\Phi(x)$. $\Phi=+\infty$ at $H
    x=\pi/2$. Right panel: the function $f(x)$.}
   \label{fig:F2}
\end{figure}

In this paper, we consider an analog de Sitter universe in our setup
of the QH system.  The left panel of Fig. \ref{fig:penrose1} depicts a
setup of our QH experiment with the expanding edge of the QH
system: the edge system is composed of an input static region I
($L/2<x$), an expanding region II ($-L/2\le x\le L/2$), and a output
static region III ($x<-L/2$).  The analog metric of this system is
written as
%%%
\begin{align}
  &ds^2=a^2(t)(-dt^2+dx^2),\notag \\
  & a(t)=\begin{cases}
           \dfrac{1}{\cos(H
           t)}\theta(t)+\theta(-t)&\quad\text{for}\quad-L/2\le x\le
                                    L/2\quad\text{(region II)},\\
           1&\quad\text{for}\quad L/2<|x|\quad\text{(region I,III)}.
           \end{cases}
  \label{eq:metric}
\end{align}
%%%
where $\theta(t)$ is the Heaviside function.  Using the proper time
$\tau=\int_0^t dt'a(t')$,  the metric in region II is
%%%
\begin{equation}
  ds^2
  =\begin{cases}
     -d\tau^2+\cosh^2(H\tau)dx^2,&\quad\text{for}\quad \tau\ge0.\\
     -d\tau^2+dx^2, &\quad\text{for}\quad \tau<0.
   \end{cases}
 \end{equation}
 %%% 
 Thus, we assume a spacetime that is flat Minkowski for
 $t<0$ and de Sitter expansion starts at $t=0$ in region II. The global structure
 of this spacetime is shown in the right panel of
 Fig. \ref{fig:penrose1} with the parameter $\pi/4<LH<\pi/2$. There
 emerges formation of  the future de Sitter horizon $\mathcal{H}^+$ in region II.
 %%% 
%%%
It is possible to obtain a relation between spatial coordinates of
regions I and II explicitly. Null coordinates in regions I and III are
related by the following formula (see detail in Appendix A):
%%%
\begin{align}
  &x_\text{I}^+=\Phi[-L+\Phi^{-1}[x_\text{III}^++\Phi[L/2]]]+L/2=:
    f(x_\text{III}^+),\\
  &x_\text{III}^-=\Phi[-L+\Phi^{-1}[x_\text{I}^{-}+L/2]]-\Phi[L/2]+L.
\end{align}
%%%
where the function $\Phi$ is defined by (Fig. \ref{fig:F2})
%%%
\begin{align}
  \Phi(x)&=\int_0^x dy\, a(y) \notag \\
         &=\begin{cases}
             \dfrac{1}{2H}\ln\dfrac{1+\sin Hx}{1-\sin Hx} &\quad
             \text{for}\quad 0\le x<\dfrac{\pi}{2H}, \\
            x&\quad\text{for}\quad x<0. 
           \end{cases}
\end{align}
%%%
%%%
\noindent
The inverse function is
%%%
\begin{equation}
  \Phi^{-1}(x)=
  \begin{cases}
    \dfrac{1}{H}\,\mathrm{arcsin}\tanh(Hx)&\quad\text{for}\quad x>0 \\
    x&\quad\text{for}\quad x<0
  \end{cases},
\end{equation}
%%%
and for $x\rightarrow+\infty$,
%%%
\begin{equation}
  \Phi^{-1}(x)\sim\frac{\pi}{2H}-\frac{2}{H}e^{-Hx}.
\end{equation}
%%%
The asymptotic form of the function $f(x^+_\text{III})=x^+_\text{I}[x^+_\text{III}]$ is
%%%
\begin{equation}
 f(x_\text{III}^+)\sim
  \begin{cases}
    c_0-c_1 e^{-Hx_\text{III}^+}&\quad\text{for}\quad
    x_\text{III}^+\rightarrow+\infty \\
    x_\text{III}^+&\quad\text{for}\quad x_\text{III}^+\rightarrow-\infty
  \end{cases}
  \label{eq:x1}
\end{equation}
%%%
where the constants $c_0$ and $c_1$ are
%%%
\begin{equation}
  c_0=\frac{L}{2}-\ln\tan(HL/2),\quad c_1=\frac{2}{H\sin
    HL}\sqrt{\frac{1-\sin HL/2}{1+\sin HL/2}}.
\end{equation}
%%%

%%%%%%%%%%%%%%%%%%%%%%%%%
%%%%%%%%%%%%%%%%%%%%%%%%%%%%%%%%%%%%%%%%%%%%%%%%%%%%%%%%%%%%%%%%%%%%%
\section{Classical simulation of Hawking radiation}
\begin{figure}[H]
  \centering 
  \includegraphics[width=0.9\linewidth,clip]{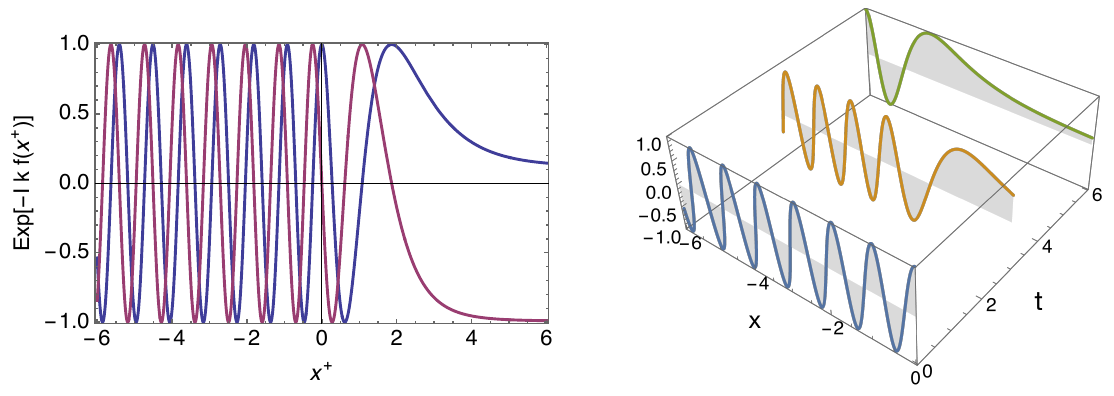}
  \caption{Wave forms in region III ($L=H=1, k=7$). Left panel : real
    part (blue) and imaginary part (red) of $\varphi$. Right panel:
    change of the spatial profile of waves (real part of $\varphi$) at different
    times ($t=0,3,6$).}
  \label{fig:F3}
\end{figure}

We send  plane waves from region I (in-region) and detect them at a point in region
III (out-region). Normalized wave modes in regions I and III are
%%%
\begin{equation}
  \varphi_k^\text{(I)}=\frac{e^{-ikx_\text{I}^+}}{\sqrt{4\pi
      k}}=\frac{e^{-ikf(x_\text{III}^+)}}{\sqrt{4\pi k}},\quad
  \varphi_k^\text{(III)}=\frac{e^{-ikx_\text{III}^+}}{\sqrt{4\pi
      k}},\quad k>0.
\end{equation}
%%%
An input plane wave $e^{-ikx^+_\text{I}}$ in region I has the wave
form $\exp(-ikf(x^+_\text{III}))$ in region III. Distortion of plane
waves due to de Sitter expansion of region II is encoded in the
function $f(x_\text{III}^+)$. Figure \ref{fig:F3} depicts wave forms
in region III. The left panel shows the real part and the imaginary
part of $\varphi_k^{(\text{I})}$ as the function of
$x^+_{\text{III}}$. The right panel shows snapshots of wave forms
$\mathrm{Re} ~\varphi_k^{(\text{I})}$ at $t=0, 3, 6$. We can observe
the wave is stretched by de Sitter expansion in region II.
%%%
%%%
\noindent
The in-mode and out-mode are related by the Bogoliubov transformation,
%%%
\begin{align}
  &\varphi_k^\text{(I)}=\int_0^\infty
    dk'\left[\al(k,k')\,\varphi_{k'}^\text{(III)}+\beta(k,k')\,
    \varphi_{k'}^\text{(III)*}\right], \\
  &\varphi_k^\text{(III)}=\int_0^\infty
    dk'\left[\al^*(k',k)\,\varphi_{k'}^\text{(I)}-\beta(k',k)\,
    \varphi_{k'}^\text{(I)*}\right],
\end{align}
%%%
%%%
where Bogoliubov coefficients $\al$ and $\beta$ are obtained from the relation
%%%
\begin{equation}
  \frac{e^{-ik'f(x_\text{III}^+)}}{\sqrt{k'}}=\int_0^\infty
  \frac{dk}{\sqrt{k}}\left[\al(k,k')\,e^{-ik x_\text{III}^+}+\beta(k,k')\,e^{ikx_\text{III}^+}\right].
\end{equation}
%%%
Thus,
%%%
\begin{equation}
  \al(k,k')=\frac{1}{2\pi}
  \sqrt{\frac{k}{k'}}\int^{\infty}_{-\infty}dy\,e^{-ik'f(y)}\,e^{iky},\quad
  \beta(k,k')=\frac{1}{2\pi}
  \sqrt{\frac{k}{k'}}\int^{\infty}_{-\infty}dy\,e^{-ik'f(y)}\,e^{-iky}.
\end{equation}
%%%
Using the asymptotic form \eqref{eq:x1} of $f(x)$, we have \cite{Hotta2022b}
%%%
\begin{equation}
  |\beta(k,k')|^2\sim
  \begin{cases}
    0&\quad\text{for}\quad x_\text{III}^+\rightarrow -\infty, \\
    \dfrac{1}{2\pi Hk'}\dfrac{1}{\exp(2\pi k/H)-1}&\quad\text{for}\quad
     x_\text{III}^+\rightarrow +\infty.
  \end{cases}
\end{equation}
%%%
For $x_\text{III}^+\rightarrow +\infty$, the Bogoliubov
coefficient $\beta$ shows the Planckian  distribution with a
temperature
%%%
\begin{equation}
  T_H=\frac{H}{2\pi}.
\end{equation}
%%%
This temperature coincides with the Gibbons-Hawking temperature in the
de Sitter spacetime. Thus, it is possible to detect the classical
counterpart of Hawking radiation from the cosmological horizon in a de
Sitter universe by measuring the Fourier component of wave signals
 in the Minkowski region III.

%%%%%%%%%%%%%%%%%%%%%%%%%%%%%%%%%%%%%%%%%%%%%%%%%%%%%%%%%%%%%%%%%%%%%
\section{Quantum simulation of Hawking radiation }
 
From now on, we consider a quantum scalar field $\hat\varphi$ in the
expanding edge system. Our main purpose is to investigate quantum
effects of the edge mode in the quantum Hall system, which is 
measurable through the local charge density $\pa_{x^+}\hat\varphi$ of
the edge excitation.

%%%%%%%%%%%%%%%%%%%%%%%%%%%%%%%%%%%%%%%%%
\subsection{Correlation functions }
Let us consider the setup shown in Fig.~\ref{fig:penrose1}. We prepare
measurement points A and B in region III. At these points, a part of
the signals emitted from $\mathcal{I}^-$ of region I cannot reach
region III after the formation of the future de Sitter horizon
$\mathcal{H}^+$ in region II. Thus, the spacetime structure is similar
to that with a black hole formation by gravitational collapse.  We
consider detection of quantum fluctuations of the scalar field in
region III by imposing the vacuum condition at $\mathcal{I}^-$ in
region I:
%%%
\begin{equation}
  \hat a_k\ket{0_\text{I}}=0.
\end{equation}
%%%
Owing to the chirality of the edge mode, we only consider a left moving
 scalar field and  the field operator in region III is expressed as
%%%
\begin{equation}
  \hat\varphi(x^{+})=\int_0^\infty\frac{dk}{\sqrt{4\pi k}}\left[\hat
    a_k\,e^{-ik f(x^+)}+\hat
    a_k^\dag\,e^{ik f(x^+)}\right].
\end{equation}
%%%
In our setup, the gauge invariant physical quantity is charge density,
which is given by the derivative of the field operator $\hat\varphi$:
%%%
\begin{equation}
  \hat\Pi(x^+):=\hat\varphi'(x^+)=-if'(x^+)\int_0^\infty
  dk\sqrt{\frac{k}{4\pi}}
  \left[\hat
    a_k\,e^{-ik f(x^+)}-\hat
    a_k^\dag\,e^{ik f(x^+)}\right].
\end{equation}
%%%
We investigate quantum effects based on the field operator $\hat\Pi$. 
Commutators between field operators are
%%%
\begin{align}
  &[\hat\varphi(x^+),\hat\varphi(y^+)]=-\frac{i}{4}\mathrm{sign}(f(x^+)-f(y^+)),\\
  &[\hat\varphi(x^+),\hat\Pi(y^+)]=\frac{i}{2}f'(y^+)\del(f(x^+)-f(y^+)),\\
  &[\hat\Pi(x^+),\hat\Pi(y^+)]=\frac{i}{2}f'(x^+)f'(y^+)\del'(f(x^+)-f(y^+)).
\end{align}
%%%
%%%
The Wightman function for $\hat\varphi$ is
%%%
\begin{align}
  D(x_1^+,x_2^+)&=\expval{\hat\varphi(x_1^+)\hat\varphi(x_2^+)} \notag
  \\
  &=\frac{1}{4\pi}\int_\mu^\infty
                  \frac{dk}{k}e^{-ik(f(x_1^+)-f(x_2^+)-i\Delta f)} \notag \\
  &=-\frac{1}{4\pi}\log[\mu(f(x_1^+)-f(x_2^+)-i\Delta f)],
\end{align}
%%%
where we introduced an IR cutoff $\mu$ as the lower bound of the integral,
and a UV cutoff $\Delta f>0$ by
%%%
\begin{equation}
  \Delta f:=f'\left((x_1^++x^+_2)/2\right)|_{x_1^+=x_2^+}\,\ep,
\end{equation}
%%%
with the spatial cutoff length $\ep$ in the flat region III.  In the
Minkowski phase $t<0$, $\Delta f=\ep$, and in the de Sitter phase
$t\ge 0$, $\Delta f\sim e^{-Ht}\ep$ which corresponds to the comoving
wavelength in the de Sitter region II. The local spatial modes
prepared in the Minkowski region III can detect long wavelength
quantum fluctuation in the de Sitter region II. In our analysis, the
scalar field $\hat\varphi$ is an effective one and there exists the
short-distance cutoff length $\ep$ below which effective treatment of
the edge mode breaks down. In the QH systems, this scale corresponds to
the magnetic length $\ell_B$.  We regard the short-distance cutoff $\ep$
as this length in our analysis.

The Wightman function for $\hat\Pi$ is
%%%
\begin{align}
  D_\Pi(x_1^+,x_2^+)
  &:=\expval{\hat\Pi(x^+_1)\hat\Pi(x^+_2)}=\pa_{x_1^+}\pa_{x_2^+}D(x_1^+,x_2^+) \notag \\
  &= \frac{f'(x^+_1)f'(x^+_2)}{4\pi}\int_0^\infty dk
    k\,e^{-ik(f(x_1^+)-f(x_2^+)-i\Delta f)}
    \notag \\
  &=-\frac{1}{4\pi}\frac{f'(x_1^+)\,f'(x_2^+)}
    {\left(f(x_1^+)-f(x_2^+)-i\Delta f\right)^2}.
\end{align}
%%%
This quantity is independent of the IR cutoff $\mu$.  Using
\eqref{eq:x1}, the asymptotic behavior becomes
%%%
\begin{equation}
D_\Pi(x_{1}^+,x_{2}^+)\sim -\frac{1}{4\pi}
  \begin{cases}
    \dfrac{1}{\left[(2/H)\sinh(H(x_{1}^+-x_{2}^+)/2)-i\ep\right]^2}&\quad\text{for}\quad
      x_{{1,2}}^+\rightarrow+\infty\\
    \dfrac{1}{(x_{1}^+-x_{2}^+-i\ep)^2}&\quad\text{for}\quad x_{{1,2}}^+\rightarrow-\infty
  \end{cases}
\end{equation}
%%%
For $x_1^+, x_2^+\rightarrow+\infty$, $D_\Pi$ has the same behavior
as that of a thermal state with the Gibbons-Hawking temperature $T_H$.
%%%
Correlation functions are
%%%
\begin{align}
  &\expval{\left\{\hat\varphi(x^+),\hat\varphi(y^+)\right\}}=\frac{1}{2\pi}\int_\mu^\infty
    \frac{dk}{k}\cos(k(f(x^+)-f(y^+)))e^{-k\Delta f},\\
  &\expval{\left\{\hat\varphi(x^+),\hat\Pi(y^+)\right\}}=\frac{f'(y^+)}{2\pi}\int_0^\infty
    dk\sin(k(f(x^+)-f(y^+)))e^{-k\Delta f},\\
  &\expval{\left\{\hat\Pi(x^+),\hat\Pi(y^+)\right\}}=\frac{f'(x^+)f'(y^+)}{2\pi}\int_0^\infty
    dk k\cos(k(f(x^+)-f(y^+)))e^{-k\Delta f}.
\end{align}

%%%%%%%%%%%%%%%%%%%%%%%%%%%%%
\subsection{Correlation of local spatial  mode}
We consider the measurement of the field $\hat\Pi(x^+)$ at $x_\text{A}$
and $x_\text{B}$ in region III. This measurement process can be represented by
the  interaction between the field operator $\hat\Pi$ and the canonical
variables $(\hat Q_D,\hat P_D)$ of the measurement apparatus. In the
present analysis, we do not  specify details of the
apparatus. The interaction Hamiltonian between the field operator and
the apparatus is
%%%
\begin{equation}
  H_\text{int}=\sum_{j=\text{A,B}}\lambda_j(t)g_j(\hat Q_D,\hat P_D)\otimes\int dx\, w_j(x)\hat\Pi(t+x),
\end{equation}
%%%
where $g_j(\hat Q_D,\hat P_D)$ is a function of canonical variables of the
measurement apparatus, $w_j(x)$ is a  window function defining a
spatial local mode of the field at $x_\text{A,B}$, and $\lambda_j(t)$ is a
switching function of the interaction. After acting on the apparatus state, this
interaction causes a change of the ``reading'' of the apparatus
depending on the state of the quantum field $\hat\Pi$ at $x_\text{A,B}$. In the
present analysis, we do not introduce details of measurement protocols
but just pay attention to the behavior of the local mode of the
quantum field introduced by
the spatial window function $w_\text{A,B}(x)$.

\begin{figure}[H]
  \centering
  \includegraphics[width=0.45\linewidth]{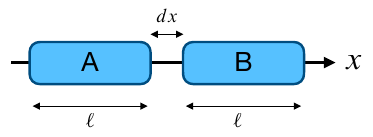}
  \caption{Setup of defining spatial regions A and B. The center of
    each region is $x_\text{A}=-L/2-3\ell/2-dx$ and
    $x_\text{B}=-L/2-\ell/2$. $x_B-x_A=\ell+dx$. }
    \label{fig:setup}
\end{figure}

\begin{figure}[H]
  \centering
  \includegraphics[width=0.7\linewidth]{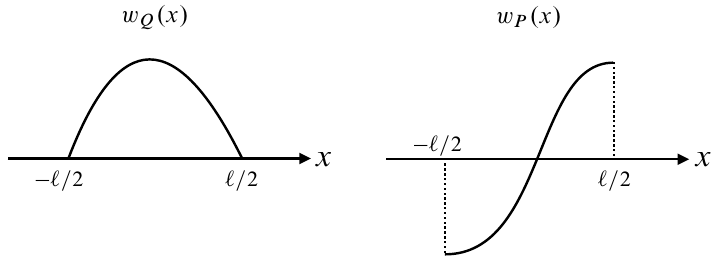}
  \caption{Spatial profile of adopted window functions.}
    \label{fig:F5}
\end{figure}

For the purpose of observing spatial correlations of the field, we
define a canonical pair of variables corresponding to the local
spatial mode of the field at $x_{A}$ and $x_{B}$:
%%%
\begin{equation}
  \hat Q_j(t)=\int dx\, w_Q(x-x_j)\hat\Pi(t+x),\quad   \hat P_j(t)=\int
  dx\, w_P(x-x_j)\hat\Pi(t+x),\quad j=\text{A,B}.
\end{equation}
%%%
We assume the window functions $w_{P,Q}(x)$ have nonzero values only in a compact spatial region
$x\in[-\ell/2,\ell/2]$.  Requiring these variables to be  canonical
pairs, equal time commutators between these variables should be
%%%
\begin{align}
  &[\hat Q_j, \hat P_k]=\frac{i}{2}\int dx\, w_Q(x-x_j)w_P'(x-x_k)\equiv
    i\del_{jk}, \quad j,k=\text{A,B},
    \label{eq:s-mode1}\\
  &[\hat Q_j, \hat Q_k]= \frac{i}{2}\int dx\,
    w_Q(x-x_j)w_Q'(x-x_k)\equiv 0,\\
  &[\hat P_j, \hat P_k]=\frac{i}{2}\int dx\,
    w_P(x-x_j)w_P'(x-x_k)\equiv 0.
\end{align}
%%%
These equations provide conditions for window functions and are
independent of the state of the quantum field.  Thus, the local
spatial modes $(\hat Q_\text{A},\hat P_\text{A})$,
$(\hat Q_\text{B},\hat P_\text{B})$ associated with spatial regions A
and B can be introduced by using suitably chosen window functions
$w_Q(x)$ and $w_P(x)$ irrespective of the states of the quantum field
(Fig. \ref{fig:setup}). Regions A and B are assumed to have no overlap
and their separation is $dx$. Locality of the spatial modes is guaranteed if
we adopt window functions with compact support. The center of each
region is assumed to be
%%%
\begin{equation}
  x_\text{A}=-\frac{L}{2}-\frac{3\ell}{2}-dx,\quad
    x_\text{B}=-\frac{L}{2}-\frac{\ell}{2},\quad x_\text{B}-x_\text{A}=\ell+dx.
\end{equation}
    %%% 
%%%
\noindent
We choose the
following window functions in our analysis (Fig. \ref{fig:F5}):
\begin{equation}
  w_Q(x)=\frac{2}{\sqrt{\pi}}\cos\left(\frac{\pi x}{\ell}\right),\quad
  w_P(x)=\frac{2}{\sqrt{\pi}}\sin\left(\frac{\pi x}{\ell}\right),\quad x\in[-\ell/2,\ell/2].
  \label{eq:window}
\end{equation}
%%%
%%%
\noindent
These window functions satisfy the condition \eqref{eq:s-mode1} and
the window functions defining the bipartite state for canonical
variables
$(\hat Q_\text{A},\hat P_\text{A},\hat Q_\text{B},\hat P_\text{B})$ do
not have spatial overlap for $x_\text{B}-x_\text{A}\ge\ell$. Equal
time correlations of these canonical variables are
%%%
\begin{align}
  &\langle\hat Q_\text{A}\hat Q_\text{B}+\hat Q_\text{B}\hat
    Q_\text{A}\rangle=
      \int dx dy\, w_Q(x-x_\text{A})w_Q(y-x_\text{B})\expval{\left\{\hat\Pi(t+x),\hat\Pi(t+y)\right\}},\\
  &\langle\hat P_\text{A}\hat P_\text{B}+\hat P_\text{B}\hat
    P_\text{A}\rangle=
    \int dx dy\, w_P(x-x_\text{A})w_P(y-x_\text{B})\expval{\left\{\hat\Pi(t+x),\hat\Pi(t+y)\right\}},\\
    &\langle\hat Q_\text{A}\hat P_\text{B}+\hat P_\text{B}\hat
    Q_\text{A}\rangle=
    \int dx dy\, w_Q(x-x_\text{A})w_P(y-x_\text{B})\expval{\left\{\hat\Pi(t+x),\hat\Pi(t+y)\right\}}.
\end{align}
%%%
As the bipartite state $\rho_\text{AB}$ defined by these canonical
variables is Gaussian, the state is determined by the covariance
matrix
%%%
\begin{equation}
  V_\text{AB}=
  \begin{bmatrix}
    a_1&a_3&c_1&c_3\\
    a_3&a_2&c_4&c_2\\
    c_1&c_4&b_1&b_3\\
    c_3&c_2&b_3&b_2
  \end{bmatrix},
\end{equation}
%%%
where its components are defined by
%%%
\begin{align}
  c_1&=\frac{1}{2}\langle\hat Q_\text{A}\hat Q_\text{B}+\hat Q_\text{B}\hat
    Q_\text{A}\rangle, \quad
  c_2=\frac{1}{2}\langle\hat P_\text{A}\hat P_\text{B}+\hat P_\text{B}\hat
    P_\text{A}\rangle, \quad
    c_3=\frac{1}{2}\langle\hat Q_\text{A}\hat P_\text{B}+\hat P_\text{B}\hat
    Q_\text{A}\rangle, \quad
    c_4=\frac{1}{2}\langle\hat Q_\text{B}\hat P_\text{A}+\hat P_\text{A}\hat
    Q_\text{B}\rangle, \\
    a_1&=\langle\hat Q^2_\text{A}\rangle,\quad
    a_2=\langle\hat P^2_\text{A}\rangle,\quad
  a_3=\frac{1}{2}\langle\hat Q_\text{A}\hat P_\text{A}+\hat P_\text{A}\hat
    Q_\text{A}\rangle,\quad b_j=a_j(\text{A}\rightarrow \text{B}).
\end{align}
%%% 

\begin{figure}[H]
  \centering
  \includegraphics[width=1.01\linewidth]{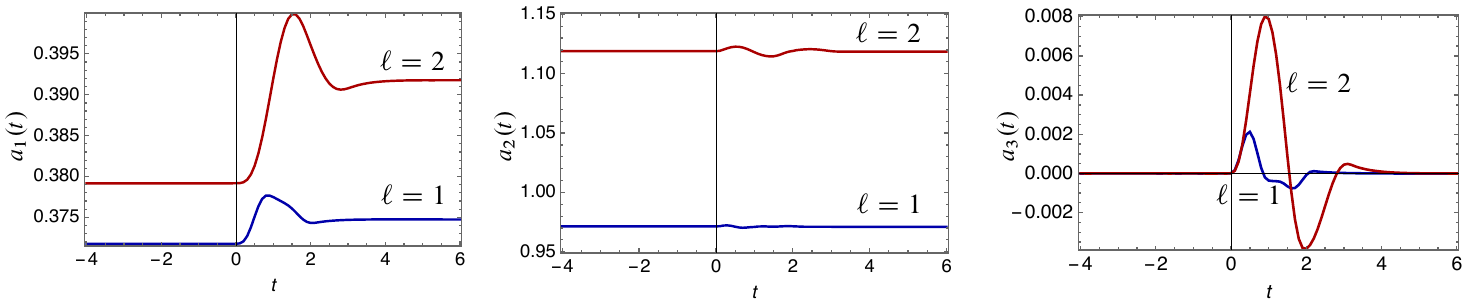}
  \caption{Behavior of the autocorrelation
    functions with different region size $\ell=1$ (blue lines) and $\ell=2$ (red lines). ($H=L=1,\ep=0.01$)}
  \label{fig:ac1}
\end{figure}

We first show temporal behavior of autocorrelation functions of the local
spatial mode in region III.  The behavior of the autocorrelation
functions $a_{1,2,3}(t)$ with different region size $\ell=1,2$ are shown
in Fig. \ref{fig:ac1}. We can observe a signature of de Sitter
expansion in region II as a change of correlations around $0<t<2$.
%%%
%%%%%
\noindent
These quantities are measurable as output signals of the detector in
our QH experiment.  To obtain a qualitative understanding of the
behavior of the autocorrelation functions, we evaluate
$a_1(t)=\langle\hat Q_\text{A}^2\rangle$ analytically with a window
function $w_Q(x)=w_0\,\theta(\ell/2+x)\theta(\ell/2-x)$ where $w_0$ is
a normalization constant the value of which is unspecified:
%%%
\begin{align}
\langle\hat Q_\text{A}^2\rangle&=\frac{1}{4\pi}\int_0^\infty dk k\left|\int dx \,w_Q(x)f'(t+x_\text{A}+x)e^{ikf(t+x_\text{A}+x)}\right|^2e^{-\Delta f \,k} \notag \\
&=\frac{2w_0^2}{\pi}\int_0^\infty\frac{dk}{k}\sin^2\left[\frac{k}{2}\left(f(t+x_\text{A}+\ell/2)-f(t+x_\text{A}-\ell/2)\right)\right]e^{-\Delta f\,k} \notag\\
&=\frac{w_0^2}{4\pi}\ln\left[1+\frac{(f(t+x_\text{A}+\ell/2)-f(t+x_\text{A}-\ell/2))^2}{(f'(t+x_\text{A})\ep)^2}\right].
\label{eq:QA2}
\end{align}
%%%
Using the asymptotic form \eqref{eq:x1} of $f(x)$,
%%%
\begin{equation}
\langle\hat Q_\text{A}^2\rangle\sim
\begin{cases}
\dfrac{w_0^2}{4\pi}\ln\left[1+\left(\dfrac{\ell}{\ep}\right)^2\right]\approx\dfrac{w_0^2}{2\pi}\ln\left(\dfrac{\ell}{\ep}\right)
&\quad\text{for}\quad
t\rightarrow-\infty,\\
\dfrac{w_0^2}{4\pi}\ln\left[1+\left(\dfrac{\ell}{\ep}\right)^2\left(\dfrac{\sinh(H\ell/2)}{(H\ell/2)}\right)^2\right]\approx
\dfrac{w_0^2}{2\pi}\ln\left(\dfrac{\ell}{\ep}\right)+\dfrac{w_0^2}{2\pi}\ln\left(\dfrac{\sinh(H\ell/2)}{(H\ell/2)}\right)&\quad\text{for}\quad
t\rightarrow+\infty,
\end{cases}
\label{eq:c1-asym}
\end{equation}
%%%
where we assume that the region size of the detection region is far larger than the UV cutoff
and $\ell/\ep\gg 1$.  The difference of the
autocorrelation between $t=-\infty$ and $t=+\infty$ is
%%%
\begin{equation}
a_1(+\infty)-a_1(-\infty)\sim \frac{w_0^2}{2\pi}\ln\left[\frac{\sinh(H\ell/2)}{(H\ell/2)}\right].
\label{eq:c1diff}
\end{equation}
%%%
This quantity is independent of the UV cutoff and its amount depends
only on $H\ell$. Hence we expect that it reflects the signature of Hawking
radiation from the de Sitter region. For further understanding of
the behavior of $a_1=\langle\hat Q_\text{A}^2\rangle$ we pay attention to
its $\ell$ dependence. We expand $a_1(\ell)$ as
%%%
\begin{equation}
a_1(\ell)=\int_0^\infty dK\, \tilde a_1(K)e^{iK\ell}.
\end{equation}
%%%
Then, the power spectrum for $a_1$ is obtained as
%%%
\begin{equation}
P(K)=K|\tilde a_1(K)|,
\end{equation}
%%%
which represents the power of detected signals with  wave number
$K=2\pi/\ell$ corresponding to the size $\ell$ of the detection
region. We see that the power spectrum shows the Planckian
distribution with the temperature $T_H=H/(2\pi)$. Using the asymptotic
form of $a_1$ in Eq. \eqref{eq:c1-asym}, the Fourier component of
$a_1(\ell)$ is
%%%
\begin{equation}
 \tilde a_1(K)\sim
 \begin{cases}
    -\dfrac{i}{k}\left(\gamma+\ln(K\ep)-i\pi/2\right)&\quad\text{for}\quad t\rightarrow-\infty,\\
    \dfrac{H}{2K^2}\left[1-i\dfrac{2K}{H}\left(\gamma+\psi(-iK/H)\right)\right]&\quad\text{for}\quad t\rightarrow+\infty,
 \end{cases}
\end{equation}
%%%
where $\gamma$ is Euler's constant and $\psi(x)=\Gamma'(x)/\Gamma(x)$ is the poly Gamma function. For $t\rightarrow-\infty$, the power spectrum of the signal is
%%%
\begin{equation}
 P(K)\sim-\ln(K\ep).
 \label{eq:pow1}
\end{equation}
%%%
For $t\rightarrow+\infty$, the power spectrum of the signal is 
%%%
\begin{align}
 P(K)&\sim 
 \dfrac{\pi}{e^{2\pi K/H}-1}\quad
 \text{for}\quad K<H.
 \label{eq:pow2}
\end{align}
%%%
Therefore, for long wavelength modes larger than the de Sitter horizon
size $H^{-1}$, the power spectrum observed in region III shows the
Planckian distribution with temperature $T_H=H/(2\pi)$ originated from
Hawking radiation in the de Sitter region II. Comparing
\eqref{eq:pow1} and \eqref{eq:pow2}, the power $P(t=+\infty)$ is
larger than $P(t=-\infty)$ in the long wavelength region $K<H$ due to
the Hawking radiation. This enhancement or amplification of the power
for long wavelength fluctuations larger than the de Sitter horizon
$H^{-1}$ has the same physical origin as that of the generation of
primordial quantum fluctuations in cosmic inflation.

\begin{figure}[H]
  \centering
    \includegraphics[width=0.5\linewidth]{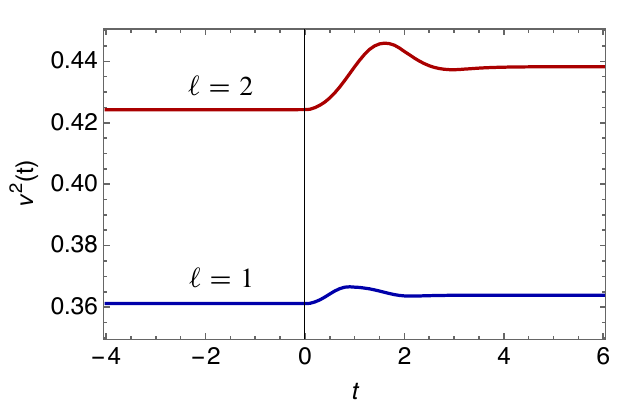}
    \caption{Evolution of the symplectic eigenvalue of the state
      $\rho_\text{A}$. $\nu^2-1/4$ represents the mixedness of this state,
      which also represents the amount of entanglement between A and $\overline{\text{A}}$.}
    \label{fig:nu2}
\end{figure}

To understand the behavior of Hawking radiation in more detail, we
consider the covariance matrix of a single mode
$(\hat Q_\text{A},\hat P_\text{A})$,
%%%
\begin{equation}
  V_\text{A}=
  \begin{bmatrix}
    a_1&a_3\\ a_3&a_2
  \end{bmatrix},
\end{equation}
%%%%%
and determinant of this matrix, which is the square of the symplectic
eigenvalue $\nu^2$ of the reduced state $\rho_\text{A}$. The physical
condition of the state requires $\nu^2\ge1/4$ and $\nu^2=1/4$
corresponds to a pure state. As we are considering a single subregion
A of the entire space, the state $\rho_\text{A}$ is mixed and the
mixedness represents the amount of entanglement between A and its
complement $\overline{\text{A}}$. Temporal behavior of $\nu^2$
(Fig. \ref{fig:nu2}) shows Hawking radiation causes an increase of
mixedness of the mode A and enhances entanglement between A and
$\overline{\text{A}}$. The mode A and $\overline{\text{A}}$ constitutes
a pure two mode squeezed state and its amount of squeezing and
entanglement between A and $\overline{\text{A}}$ increases due to the Hawking
radiation created by the rapid accelerated expansion of the background
space.
%%%
%%%
%%%%%%%%%%%%%%%%%%%%%%%%%%%%%%%%%%%%%%%%%%%%%%%%%%%%%%%%%%
\section{Entanglement of spatial  modes} 
We investigate behavior of entanglement between spatial regions A and
B in region III (Fig. \ref{fig:setup}) using associated spatial local
modes $(\hat Q_\text{A},\hat P_\text{A})$ and
$(\hat Q_\text{B},\hat P_\text{B})$. We can evaluate entanglement
negativity from the symplectic eigenvalues of the covariance matrix
$V_\text{AB}$; the covariance matrix $V_\text{AB}$ has two symplectic
eigenvalues $\nu_{\pm}\ge 1/2$, and the partially transposed
covariance matrix $\tilde V_\text{AB}$ has two symplectic eigenvalues
$\tilde\nu_{\pm}$.  Based on the positivity criterion of the partially
transposed covariance matrix for bipartite Gaussian
states \cite{Peres1996,Horodecki1997,Simon2000}, a measure of entanglement between A and B is
given by the logarithmic negativity defined as \cite{Vidal2002a,Plenio2005}
%%%
\begin{equation}
  E_N:=-\mathrm{min}[\log_2(2\tilde\nu_{-}),0].
\end{equation}
%%%
For $E_N>0$, the bipartite state $\rho_\text{AB}$ is entangled and the
logarithmic negativity represents the amount of entanglement between A
and B.
%%%%%%%%%%%%%%%%%%%%%%%%%%%%%%%%

%%%%%%%%%%%%%%%%%%%%%%%%%%%%%%%%%%%%%%%%%%%%%%%%%%%%%%%%%
\subsection{Minkowski case}
\begin{figure}[H] 
  \centering
  \includegraphics[width=0.85\linewidth]{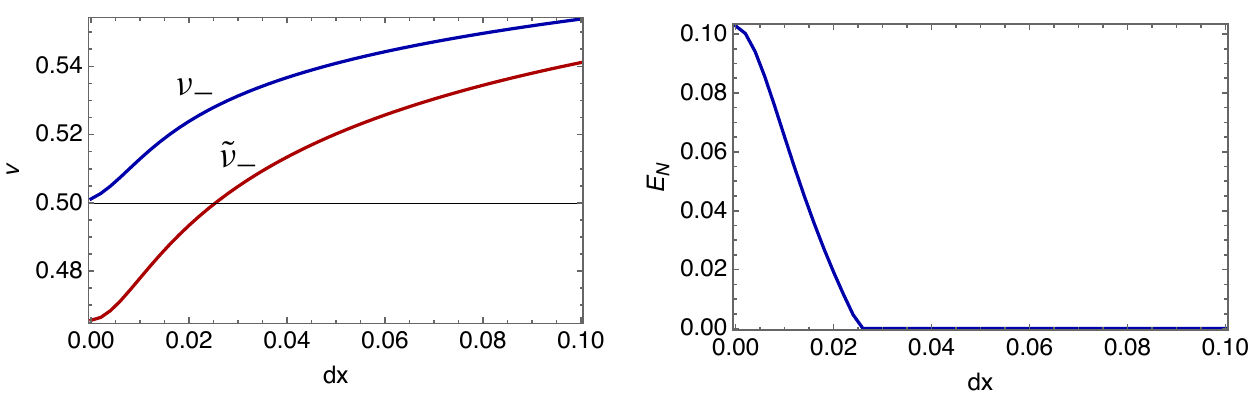}
  \caption{Left panel: separation dependence of symplectic eigenvalues
    $\nu_{-}$(blue), $\tilde\nu_{-}$(red). Positivity of the bipartite
    state $\nu_{-}\ge 1/2$ is preserved for $dx\ge 0$. For
    $\tilde\nu_{-}<1/2$, A and B are entangled. Right panel:
    separation dependence of logarithmic negativity for the Minkowski
    case. The bipartite state $\rho_\text{AB}$ becomes separable for
    large separation ($\ep=0.01, \ell=1, L=1$). The critical
    separation depends on the values of the UV cutoff $\ep$.}
  \label{fig:neg-M}
\end{figure}

\begin{figure}[H] 
  \centering
  \includegraphics[width=0.9\linewidth]{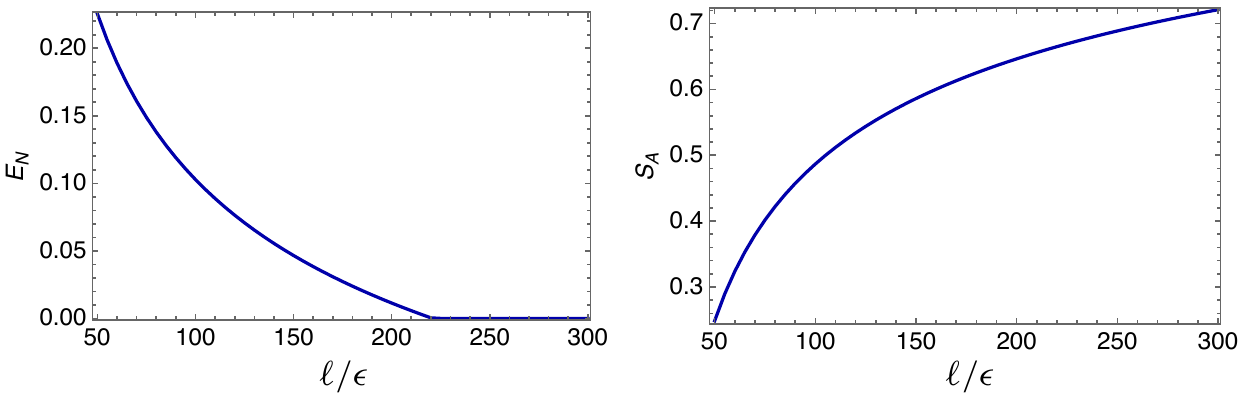}
  \caption{Left panel: $\ell/\ep$ dependence of negativity with
    $dx=0$. Right panel: $\ell/\ep$ dependence of entanglement entropy
    for a single region A. For $\ell/\ep\gg1$, the entanglement
    entropy behaves as $S_A\propto \log(\ell/\ep)$.}
  \label{fig:neg-ep}
\end{figure}

We first show the behavior of entanglement for the case that region II
is static and there is no expanding edge region (Minkowski case). We
confirm that the detection of entanglement between A and B is possible
using local modes defined by window functions \eqref{eq:window}. The
left panel of Fig. \ref{fig:neg-M} shows the symplectic eigenvalues as a
function of separation $dx$ of two regions, and the right panel of
Fig. \ref{fig:neg-M} shows the negativity as a function of $dx$. There
exists a critical separation and below this separation entanglement
between A and B can be detected.
%%%
%%%
\noindent
The left panel of Fig. \ref{fig:neg-ep} shows the dependence of the
cutoff parameter $\ell/\ep$ on negativity with $dx=0$. For large
values of the ratio $\ell/\ep$, the bipartite system AB becomes
separable and local modes cannot detect entanglement of the scalar
field.  It is possible to understand this behavior from the viewpoint
of entanglement monogamy \cite{Hiroshima2007}.  Let us focus on the
entanglement entropy of region A which is a subsystem of the entire
spatial region. The entanglement entropy for a single Gaussian mode
$(\hat Q_\text{A},\hat P_\text{A})$ is given by \cite{Holevo2001}
  %%%
  \begin{equation}
  S_\text{A}=(\nu+1/2)\log_2(\nu+1/2)-(\nu-1/2)\log_2(\nu-1/2)
  \end{equation}
  %%%
  with the symplectic eigenvalue $\nu$ of the covariance matrix
  $V_\text{A}$ for the mode A. As is shown in the right panel of
  Fig. \ref{fig:neg-ep}, $S_\text{A}$ behaves
  $\propto\log(\ell/\ep)$,\footnote{This behavior is confirmed
    numerically and we do not derive this relation analytically.}
  which is the typical scaling behavior of entanglement entropy of the
  massless scalar field in the 1+1 dimensional case
  \cite{Bombelli1986,Srednicki1993}. This behavior implies that
  entanglement between region A and its complement becomes larger as
  $\ell/\ep$ increases. Concerning entanglement between A and B,
  because A and its complement, and B and its complement become
  strongly entangled as $\ell/\ep$ becomes larger, owing to the monogamy
  property of multipartite entanglement
  \cite{Bengtsson2016,Hiroshima2007}, the entanglement between A and B
  should become smaller and the entanglement between A and B vanishes
  above a some critical value of $\ell/\ep$.
%%%%% 

%%%%%%%%%%%%%%%%%%%%%%%%%%% 
\subsection{De Sitter case}
\begin{figure}[H]
  \centering 
  \includegraphics[width=0.9\linewidth]{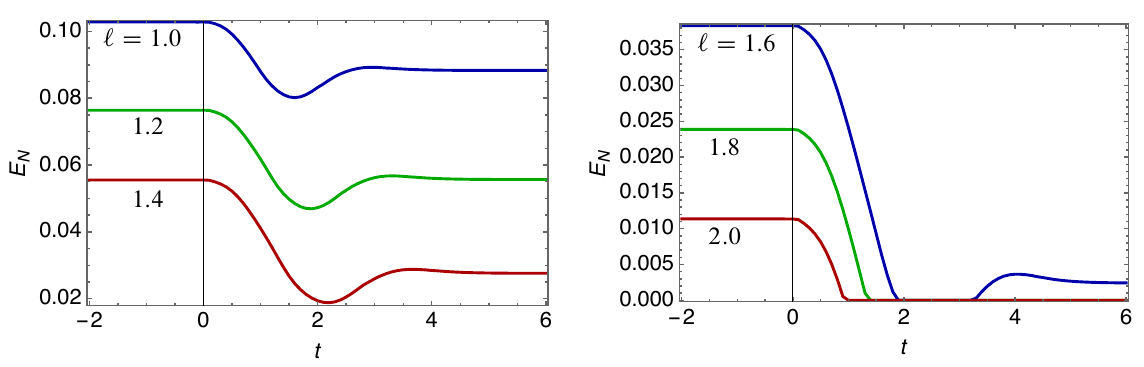}
  \caption{Evolution of negativity between regions A and B for the de
    Sitter case with different spatial region size $\ell$
    ($dx=0, \ep=0.01, H=L=1$). Because of Hawking radiation from the de
    Sitter region, negativity decreases around $t=0\sim2$. The final
    value of negativity becomes smaller than the initial negativity in
    the Minkowski region. For $\ell=1.8,2.0$, the final value of
    negativity becomes zero and entanglement death occurs. For
    $\ell=1.6$, both death and revival of entanglement are observed.}
  \label{fig:neg-D} 
\end{figure}

\begin{figure}[H]
  \centering 
  \includegraphics[width=0.5\linewidth]{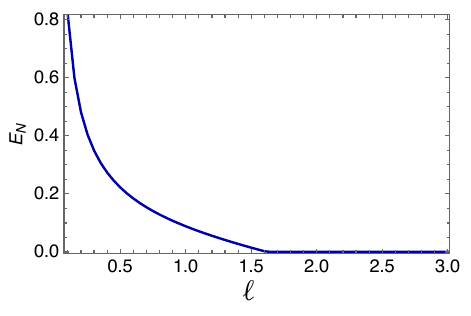}
  \caption{Region size dependence of negativity at $t=6$
    ($dx=0,L=H=1, \ep=0.01$). A and B become separable for large
    scales $\ell\ge1.65H^{-1}$.}
  \label{fig:neg-D2} 
\end{figure}

We move on to the expanding edge case which mimics a de Sitter
universe and consider entanglement between adjacent regions A and B in
region III under the influence of Hawking radiation from the de Sitter
region II.  Figure \ref{fig:neg-D} shows the evolution of entanglement
between A and B with different sizes of A and B with $dx=0$. As we can
observe from Fig. \ref{fig:neg-D}, following the transient change of
negativity during $0<t<3$, which is determined by the shape of the
window function, the negativity becomes asymptotically constant. The
final amount of entanglement is reduced compared to the initial
Minkowski value. The reduction of entanglement depends on the size of
spatial regions $\ell$. For $\ell=1.0-1.4$, a nonzero value of
negativity survives at $t=6$. On the other hand, for sufficiently
large region size $\ell\gg H^{-1}$, which corresponds to detection of
long wavelength superhorizon fluctuation in the de Sitter universe,
the entanglement between A and B becomes zero after arrival of the
Hawking radiation. This behavior of ``entanglement death'' is the same
as that confirmed in inflationary models
\cite{Nambu2008,Nambu2011,Matsumura2018} and is responsible for the
emergence of classical behavior from quantum fluctuations. Thus, using
our setup of the QH experiment, it is possible to simulate ``classical
to quantum transition'' of primordial quantum fluctuations in a
laboratory.
%%% 
%%%
Figure \ref{fig:neg-D2} shows the region size dependence of the negativity at
$t=6$. For $\ell\ge 1.65$, the negativity becomes zero and the two regions A
and B become separable. The quantum correlation between the two regions is
lost for large scales compared to the de Sitter horizon length
$H^{-1}$. For these large scales, spatial correlations between A and B
exist as classical correlations. Therefore, the long wavelength
Hawking radiation can be treated as classical stochastic fluctuations
and we can confirm the classicality of Hawking radiation originated from
zero point quantum fluctuations of the scalar field.
%%%

%%%%%%%%%%%%%%%%%%%%%%%%%%%%%%%%%%%%%%%%%%
\section{Summary and Speculation}

We considered the analog de Sitter universe realized by the expanding
edge of a QH system. We investigated the behavior of the chiral
massless scalar field corresponding to an edge excitation, and
discussed the detection of Hawking radiation from the de Sitter
region. In our setup of the expanding edge system, the spacetime
structure is similar to that of a black hole formation via
gravitational collapse; the future event horizon is formed and Hawking
radiation with thermal spectrum from the vicinity of the future event
horizon is expected. The entanglement between spatial regions A and B
in the flat region is also evaluated and we found that Hawking
radiation from the de Sitter region reduces preexistent entanglement
before the arrival of Hawking radiation, and for sufficient large size
of detection regions compared to the Hubble length in the de Sitter
region, the two regions become separable and only classical
correlation survives. This behavior is the same as that appearing in
cosmic inflation.  To conduct the UV divergence of the quantum scalar
field, we introduced a UV cutoff as the scale at which the effective
field treatment of the edge excitation breaks down. The correlation
functions of the local spatial modes also contain this cutoff
dependence and it is possible to examine the impact of the cutoff on
Hawking radiation using our experiment, which is related to the
trans-Planckian problem in black hole evaporation and cosmic
inflations.  It is important to investigate how the effective theory
for the edge excitation breaks down below the cutoff length since the
deviation from the effective theory may introduce corrections to the
massless Klein-Gordon equation adopted in this paper. Beside applying
the expanding edge of the QH system as a simulator of the quantum
cosmology, it is also possible to explore physics of fundamental
aspects of quantum mechanics and quantum field theory because this
system can provide the squeezed vacuum state by amplification of the
vacuum fluctuations in the expanding edge region. Thus, investigation
of the violation of macrorealism (the Legget-Garg inequality
\cite{Emary2014a}) with the quantum field and the realization of the
quantum energy teleportation \cite{Hotta2014a} are possible.

In this paper, the Hall edges are described by quantum field theory in
curved space in the long wavelength regime compared to the magnetic
length $\ell_{B}$.  As seen in the above analysis, the edge can be
regarded as a fixed 1+1 dimensional universe. It may be interesting to
point out a possibility that the same system can be described by
different effective theories of quantum gravity. For instance, let us
consider a static QH system confined in a circle edge. The edge is
regarded as a closed 1+1 dimensional universe, the spacetime curvature
curvature of which vanishes. Since the electrons located at the edge
are in a quantum state with position fluctuation, the edge fluctuates
quantum mechanically. This yields quantum superposition of edge
configurations with different edge lengths. In this sense, quantum
universes with different sizes are quantum mechanically
superposed. This suggests a realization of quantum gravity at the QH
edge. Though the precise model for the static quantum universe has not
yet been specified, the classical action may be given by the following
dilaton gravity model:
%%%
\begin{equation}
S=\int d^2 x \sqrt{-g(x)} \left(\Phi(x) R(x) +\frac{\Lambda}{\ell_B} \right),
\end{equation}
%%%
where $\Lambda$ is a positive constant, $\Phi(x)$ is a real scalar
field referred to as dilaton field, and $R(x)$ is the scalar curvature
of the 1+1 dimensional universe. Taking the variation of $S$ with
respect to $\Phi(x)$ yields $R(x)=0$ as the equation of motion. Thus, 
the classical action is evaluated as
%%%
\begin{equation}
S_\text{QG}=\frac{\Lambda}{\ell_B}\int d^2 x  \sqrt{-g(x)}  .
\end{equation}
%%%
By using a static configuration of $\varphi$ independent of $t$, let
us parametrize the edge in the $x$-$y$ plane as
%%%
\begin{equation}
(x,y)=\left(x,\ell_B \varphi(x) \right),
\end{equation}
%%%
where the edge fluctuation occurs in the $y$ direction. The induced
metric for the edge is given by
%%%
\begin{equation}
ds^2 =-dt^2+dx^2+dy^2=-dt^2 +h(x)dx^2,
\end{equation}
%%%
where $h(x)=1+\ell_B^2 (\partial_x \varphi(x))^2$. Then the value of $S_\text{QG}$ is computed as 
%%%
\begin{equation}
S_\text{QG}=\frac{\Lambda}{\ell_B}\int d^2x \sqrt{1+\ell_B^2 (\partial_x \varphi(x))^2}. \label{01}
\end{equation}
%%%
It may be worth noting that the above value of $S_\text{QG}$ can be reproduced by the classical action of the field theory of QH edges:
%%%
\begin{equation}
S_\text{QH}=\int d^2 x  \left( N(x) \frac{\Lambda^2}{4} +\frac{1}{N} \left( (\partial_x \varphi)^2 +\frac{1}{\ell_B^2} \right) \right),
\end{equation}
%%%
where $N(x)$ is a lapse function. When we take $N(x)=1$, $S_\text{QH}$ yields the action of the quantum field theory for the edges. By taking the variation of $S_\text{QH}$ with respect to $N(x)$, we get
%%%
\begin{equation}
N(x)=\frac{2}{\Lambda \ell_B}\sqrt{1 +\ell_B^2 (\partial_x \varphi)^2}.
\end{equation}
%%%
It turns out that substitution of the above $N(x)$ into $S_\text{QH}$
reproduces the value of $S_\text{QG}$ in (\ref{01}).  Though the
correct relation between the theories with $S_\text{QG}$ and
$S_\text{QH}$ remains vague at present, it may be interesting to
explore the correspondence and quantum gravity effective theory for
the QH edges. In this case, the circular edge corresponds to a closed
universe. The interpretation of the wave functions of the quantum
closed universe, which satisfy the Wheeler-DeWitt equation, can be
developed from the viewpoint of the many-body wave functions of the QH
systems.  As a last comment it is worth mentioning that subtle effects
of quantum gravity, which are predicted by recent holographic
framework based on DS/dS correspondence~\cite{DS1,DS2,DS3,Inf1,Inf2}
closely related to the subject of information loss paradox in black
holes, might yield different results in entanglement calculation from
our results based on quantum field theory in fixed curved spacetime
background. Though the analysis based on the holographic treatment is
out of  scope of this paper, this direction of investigation will reveal
the feature of quantum gravity and we hope the future QH experiments
will be capable of discriminating which theories are suitable to
effectively describe the detected behavior of entanglement.

\begin{acknowledgements}
  We would like to thank A. Matsumura, Y. Osawa, Y. Sugiyama and K. Yamamoto for providing
  their valuable insight on the subject. This research was supported
  in part by a Grant-in-Aid for Scientific Research, Grant No. 21H05188 (M.H.), No. 21H05182 (M.H.), No. JP19K03838 (M.H.), No. 19K03866 (Y.N.) and No. 22H05257 (Y.N.) from the Ministry of Education, Culture, Sports, Science, and Technology (MEXT), Japan.
\end{acknowledgements}

%\bibliography{000:My_projects,003:inflation,001:quantum_physics,sonic_analog}
%\bibliography{My_projects,sonic_analog,quantum_physics}
%\bibliography{cosmology,my-paper,relativity,quantum,black-hole}

%%%%%%%%%%%%%%%%%%%%%%%%%%%%%%%%%%%%%%%%%
%%%%%%%%%%%%%%%%%%%%%%%%%%%%%%%%%%%%%%%%%%%%%%%%%%%%%%%%%%%%%%%%%%%%%
\appendix
\section{Derivation of coordinate transformation}
We review coordinate transformation in \cite{Hotta2022b}. The metric is
%%%
\begin{align}
 & \text{region I: }~ L/2\le x_\text{I},\quad 
                                       ds^2=-dx_\text{I}^+dx_\text{I}^-
                                       \notag\\
  &\text{region II: }~-L/2\le x\le L/2,\quad
    ds^2=-e^{2\Theta(t)}dx^+dx^- \notag \\
  &\text{region III: }~x_\text{III}\le -L/2,\quad
    ds^2=-dx_\text{III}^+dx_\text{III}^- \notag
\end{align}
%%%
where $e^{\Theta(t)}$ is the scale factor in region II. We consider the
coordinate transformation of the form
$x^{+}=x^{+}[x^{+}_\text{I,III}], x^{-}=x^{-}[x^{-}_\text{I,III}]$,
which keeps conformal invariance.
%%%%%%%%%%%%%%%%%%%%%%%%%%%%
\begin{description}
  \item [Regions II and III] We look for coordinates
  $x^+_\text{III}[x^+], x^-_\text{III}[x^-]$
  which cover regions II and III. The matching point is
  $x^1=x^1_\text{III}=(x^+_\text{III}-x^-_\text{III})/2=-L/2$. The
  matching condition is
  %%%
\begin{equation}
  x^+_\text{III}[t-L/2]-x^-_\text{III}[t+L/2]=-L.
  \label{eq:matching}
  \end{equation}
  %%%
  By taking the derivative with respect to $t$, we obtain
  %%%
  \begin{equation}
    \frac{dx^+_\text{III}}{dx^+}[t-L/2]=\frac{dx^{-}_\text{III}}{dx^-}[t+L/2].
  \end{equation}
  %%%
The  matching of the metric between region II and III (at $x=x_\text{III}=-L/2$) is
  %%%
  \begin{equation}
    e^{2\Theta(t)}\frac{dx^+}{dx^+_\text{III}}\frac{dx^-}{dx^-_\text{III}}=1\quad\therefore\quad
    \frac{dx^+_\text{III}}{dx^+}[t-L/2]=e^{\Theta(t)}.
  \end{equation}
  %%%
  By shifting the argument of the functions,
  %%%
  \begin{equation}
    \frac{dx^+_\text{III}}{dx^+}[x^+]=e^{\Theta(x^++L/2)},
  \end{equation}
  %%%
  and we obtain
  %%%
  \begin{equation}
    x^+_\text{III}[x^+]=\int_0^{x^+}dy
    e^{\Theta(y+L/2)}=\int_{L/2}^{x^++L/2}dy\,e^{\Theta(y)}=\Phi[x^++L/2]-\Phi[L/2],
    \label{eq:xop}
  \end{equation}
  %%%
  where we have fixed the integration constant such that 
  $x^+_\text{III}\propto x^+$ for constant $\Theta$ and the function
  $\Phi$ is introduced by
  %%%
  \begin{equation}
    \Phi[x]=\int_0^x dy e^{\Theta(y)}.
  \end{equation}
  %%%
  %%%
  The coordinate function $x^-_\text{III}(x^-)$ can be derived from
  \eqref{eq:matching} as
  %%%
  \begin{align}
    x_\text{III}^-[x^-]
    &=x^+_\text{III}[x^--L]+L=\int_0^{x^--L}dy\,e^{\Theta(y+L/2)}+L
      \notag \\
    &=\Phi[x^--L/2]-\Phi[L/2]+L.
      \label{eq:xom}
  \end{align}
  %%%
  %%%
  From \eqref{eq:xop} and \eqref{eq:xom},
  %%%
  \begin{align}
    dx^+_\text{III}\,dx^-_\text{III}&=e^{\Theta(x^++L/2)}e^{\Theta(x^--L/2)}dx^+dx^-,
  \end{align}
  %%%
  and the metric in region II can be written as
  %%%
  \begin{align}
    ds^2_\text{II}&=-e^{2\Theta(t)}dx^+ dx^-  \notag \\
    &=-\exp\left[2\Theta(t)-\Theta(x^++L/2)-\Theta(x^--L/2)\right]dx^+_\text{III}\,dx^-_\text{III}.
  \end{align}
  %%%
  At the boundary $x^1=-L/2$, $x^+=t-L/2$ and $x^-=t+L/2$. Thus, the
  metric becomes the flat form
  $ds^2_\text{II}=-dx^+_\text{III}\,dx^-_\text{III}$, and can be extended to the
  flat region III using coordinates $x^\pm_\text{III}$.
  %%%
  \item [Regions I and II] The condition of the matching at $x^1=L/2$
  yields
  %%%
  \begin{equation}
    x^+_\text{I}[t+L/2]-x^-_\text{I}[t-L/2]=L.
  \end{equation}
  %%%
  The coordinate function is
  %%%
  \begin{equation}
    x_\text{I}^+[x^+]=\int_0^{x^+}dy\,e^{\Theta(y-L/2)}=\Phi[x^+-L/2]-\Phi[-L/2],
    \label{eq:xip}
  \end{equation}
  %%%
  %%%
  In a similar way, we obtain
%%%
  \begin{align}
    x_\text{I}^-[x^-]&=x^+_\text{I}[x^-+L]-L=\int_0^{x^-+L}dy\,e^{\Theta(y-L/2)}-L
                        \notag \\
                     &=\Phi[x^-+L/2]-\Phi[-L/2]-L.
                       \label{eq:xim}
  \end{align}
  %%%
   From \eqref{eq:xip} and \eqref{eq:xim},
  %%%
  \begin{align}
        dx^+_\text{I}\,dx^-_\text{I}&=e^{\Theta(x^+-L/2)}e^{\Theta(x^-+L/2)}dx^+dx^-,
  \end{align}
  %%%
  and the metric in region II becomes
  %%%
  \begin{align}
    ds^2_\text{II}&=-e^{2\Theta(t)}dx^+ dx^-  \notag \\
    &=-\exp\left[2\Theta(t)-\Theta(x^+-L/2)-\Theta(x^-+L/2)\right]dx^+_\text{I}\,dx^-_\text{I}.
  \end{align}
  %%%
  At the boundary $x^1=L/2$, $x^+=t+L/2$ and $x^-=t-L/2$. Thus, the
  metric becomes the flat form
  $ds^2_\text{II}=-dx^+_\text{I}\,dx^-_\text{I}$, and can be extended to the
  flat region III using coordinates $x^\pm_\text{I}$.
  
  %%%%%%%%%%%%%%%%%%%%%%%%%%%%%% 
  \item [Relation between $x_\text{I}$ and $x_\text{III}$]
  Using the function $\Phi$,
%%%
\begin{align}
    &x_\text{III}^+=\Phi(y^++L/2)-\Phi(L/2),\quad
      x_\text{I}^+=\Phi(y^+-L/2)-\Phi(-L/2), \\
    &x_\text{III}^-=\Phi(y^--L/2)-\Phi(L/2)+L,\quad
      x_\text{I}^-=\Phi(y^-+L/2)-\Phi(-L/2)-L.
\end{align}
By eliminating $y^\pm$, we obtain a connection formula between
$x_\text{I}$ and $x_\text{III}$:
%%%
\begin{align}
  &x_\text{I}^+=\Phi[-L+\Phi^{-1}[x_\text{III}^++\Phi[L/2]]]-\Phi[-L/2]=:f(x^+_\text{III}),\\
  &x_\text{III}^-=\Phi[-L+\Phi^{-1}[x_\text{I}^{-}+L+\Phi[-L/2]]]-\Phi[L/2]+L.
\end{align}
\end{description}
%%%%%%%%%%%%%%%%%%%%%%%%%%%%%%%%%%%%%%
%%%%%%%%%%%%%%%%%%%%%%%%%%%%%%%%%%%%%%%%%%%%%%%%%% 5
\section{De Sitter case : global chart}
\begin{figure}[H]
  \centering
  \includegraphics[width=0.45\linewidth,clip]{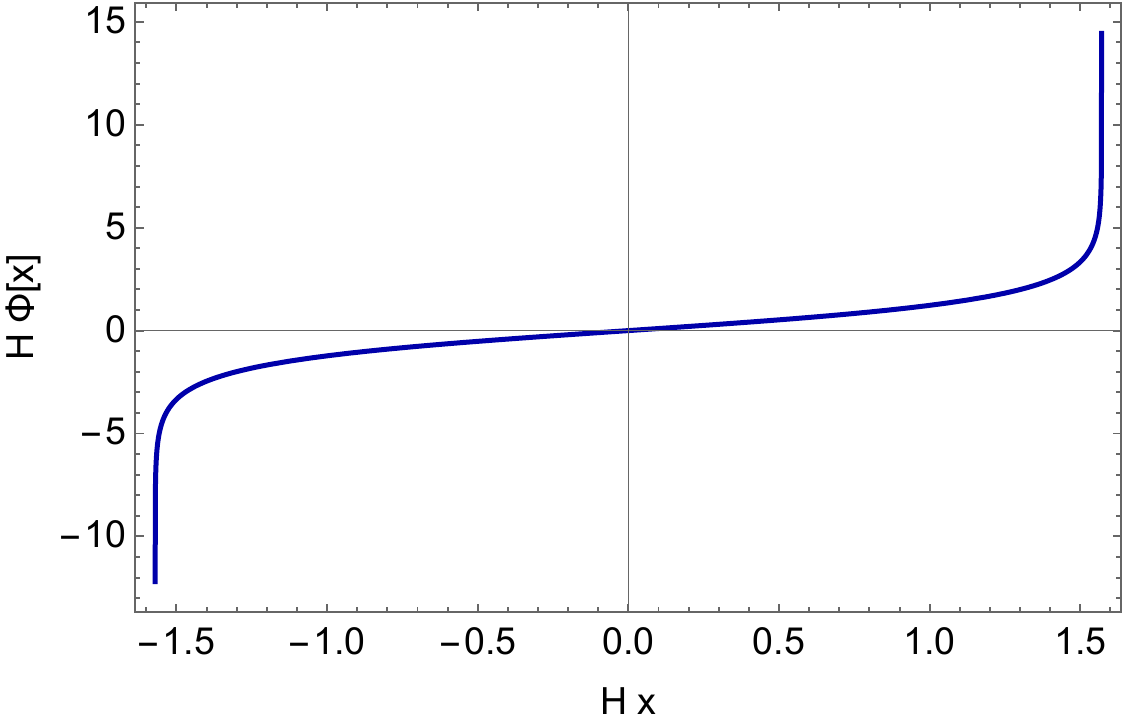}\hspace{0.5cm}
    \includegraphics[width=0.45\linewidth,clip]{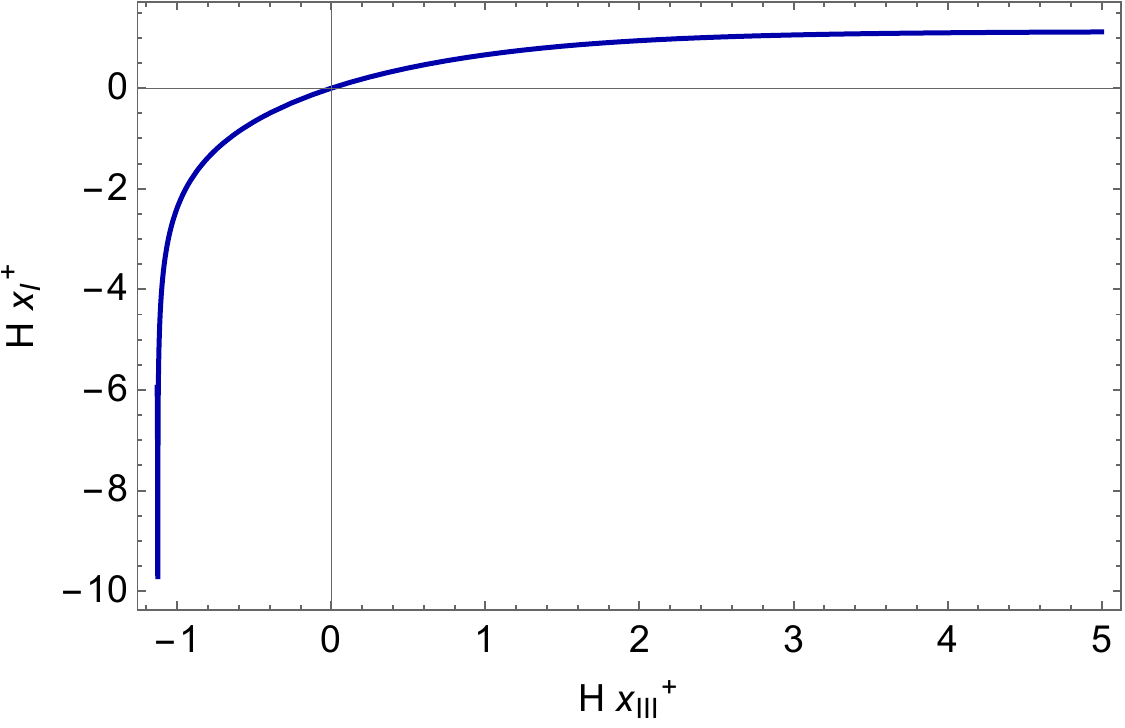}
  \caption{The function $\Phi(x)$ and
    $x_\text{I}^+=f(x^+_\text{III})$.}
  \label{fig:F12}
\end{figure}

We consider the global de Sitter spacetime. The conformal factor  is given by
%%%
\begin{equation}
  e^{\Theta(t)}=\frac{1}{\cos(Ht)},
\end{equation}
%%%
and the function $\Phi$ is (Fig. \ref{fig:F12})
%%%
\begin{equation}
  \Phi(x)=\int_0^x dy e^{\Theta(y)}=\frac{1}{2H}\ln\frac{1+\sin
    Hx}{1-\sin Hx},\quad\Phi^{-1}(x)=\frac{1}{H}\mathrm{arcsin}\tanh Hx.
\end{equation}
%%% 

\begin{figure}[H]
  \centering
  \includegraphics[width=0.45\linewidth,clip]{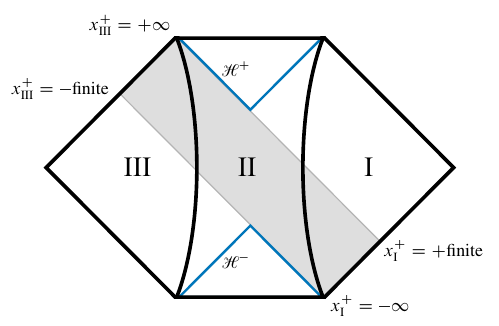}
  \caption{Penrose diagram for the global de Sitter case with $\pi/4<LH<\pi/2$.}
  \label{fig:penroseGDS}
\end{figure}

\begin{figure}[H]
  \centering
  \includegraphics[width=0.45\linewidth,clip]{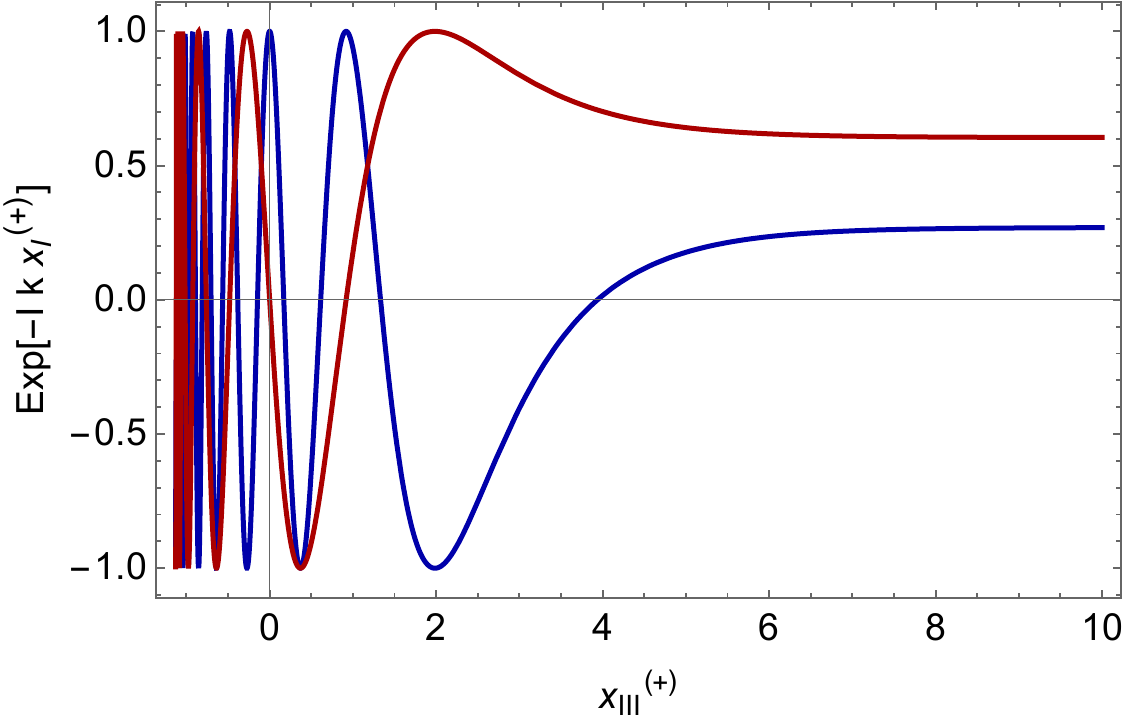}
  \caption{Wave form of $\exp(-ikf(x^+_\text{III}))$. The real
    part (blue) and the imaginary part (red) of the wave are shown.}
  \label{fig:GDS}
\end{figure}

%%%
\noindent
For $\pi/4<LH<\pi/2$, from the behavior of the function $x_\text{I}^+$, we can obtain the
global structure as shown in Fig.~\ref{fig:penroseGDS}.
%%%
%%%
Figure \ref{fig:GDS} shows a wave form of
$\exp(-ikf(x^+_\text{III}))$. As $x_\text{III}$ approaches a
finite negative value, the wavelength becomes zero, which
reflects a blueshift of waves by the past horizon $\mathcal{H}^-$. On
the other hand, for a large value of $x_\text{III}^+$, the wavelength
becomes infinite which reflects a redshift of waves by the future
horizon $\mathcal{H}^+$.
%%%%%%%%%%%%%%%%%%%

%%%%%%%%%%%%%%%%% 55

\end{document}